\documentclass[12pt]{article}
\usepackage{setspace}
\usepackage{amsmath,amssymb}
\usepackage{color}
\usepackage{hyperref}

\begin{document}
\setlength{\topmargin}{-1cm} 
\setlength{\oddsidemargin}{-0.25cm}
\setlength{\evensidemargin}{0cm}

\newcommand{\e}{\epsilon}
\newcommand{\beq}{\begin{equation}}
\newcommand{\eeq}[1]{\label{#1}\end{equation}}
\newcommand{\bea}{\begin{eqnarray}}
\newcommand{\eea}[1]{\label{#1}\end{eqnarray}}
\renewcommand{\Im}{{\rm Im}\,}
\renewcommand{\Re}{{\rm Re}\,}
\newcommand{\diag}{{\rm diag} \, }
\newcommand{\Tr}{{\rm Tr}\,}
\def\draftnote#1{{\color{red} #1}}
\def\bldraft#1{{\color{blue} #1}}
\def\n{n \cdot v}
\def\ni{n\cdot v_I}

\begin{titlepage}
\begin{center}

\vskip 4 cm

{\Large \bf  A Supertranslation-Invariant Formula for the Angular Momentum Flux in Gravitational Scattering}

\vskip 1 cm

{Reza Javadinezhad \footnote{E-mail: \href{mailto:rj1154@nyu.edu}{rj1154@nyu.edu}} and Massimo Porrati \footnote{E-mail: \href{mailto:mp9@nyu.edu}{mp9@nyu.edu}} }

\vskip .75 cm

{\em Center for Cosmology and Particle Physics, \\ Department of Physics, New York University, \\ 726 Broadway, New York, NY 10003, USA}

\end{center}

\vskip 1.25 cm

\begin{abstract}
\noindent  
The angular momentum radiated in gravitational scattering can be changed by performing a supertranslation of the asymptotic metric, 
i.e. by adding radiation with infinite wavelenght to the metric. This puzzling property can be avoided by adopting a
supertranslation-invariant definition of the angular momentum flux in general relativity. Definitions currently available in the 
literature cannot reproduce the flux necessary to obtain the correct radiation reaction effects in gravitational scattering. They also 
disagree with computations of the flux performed using scattering amplitudes and soft graviton theorems. In this paper we provide a 
new supertranslation-invariant definition of the angular momentum flux in gravitational scattering that uses only asymptotic 
metric data and reproduces the flux necessary to obtain the correct radiation reaction effects. 
\end{abstract}
\end{titlepage}
\newpage
Gravitational scattering produces radiation that carries away energy, momentum and angular momentum. These quantities can be 
defined rigorously using the formalism developed in~\cite{bond1,bms,s1,s2}. They are computed at future null infinity $\mathcal{I}^+$
 in Bondi-Sachs coordinates
\bea
 ds^2  &=&  -du^2 - 2 du\, dr + r^2 \left(h_{AB}
+\frac{C_{AB}}{r}\right) d\Theta^A d\Theta^B + D^AC_{AB}\, du\, d\Theta^B +
\frac{2m}{r} du^2  \nonumber \\ && 
		+ \frac{1}{16 r^2} C_{AB}C^{AB} du dr 
	 + \frac{1}{r}\left(
		\frac{4}{3}\left(N_A+u \partial_A m \right) - \frac{1}{8} \partial_A \left(C_{BD}C^{BD}\right)
		\right) du d\Theta^A \nonumber \\ &&
		+ \frac{1}{4} h_{AB} C_{CD}C^{CD}d \Theta^A d \Theta^B
		+ \dots , 
		\eea{1}
where the mass aspect $m(\Theta,u)$ is a scalar, the angular momentum aspect $N_A(\Theta,u)$ is a vector and the shear $C_{AB}(\Theta,u) $ is a symmetric and traceless tensor. All these quantities are defined on 
the celestial sphere with coordinates $\Theta_A$ and round metric $h_{AB}$ and also depend on the retarded time $u$. 
The dots in \eqref{1} denote subdominant terms in $1/r$. 
The coordinate system in eq.~\eqref{1} is left invariant by the asymptotic symmetries $u\rightarrow u + f(\Theta)$, called  
supertranslations~\cite{bms}. Energy and angular momentum 
are defined in terms of $m$, $N_A$ and the three Killing vectors $Y^A$ of the celestial sphere~\cite{bms}
\beq
E(u)={1\over 4\pi G} \int d^2 \Theta \sqrt{h} m(\Theta,u), \quad J_Y(u) = {1\over 8\pi G} \int d^2 \Theta \sqrt{h} Y^A N_A(\theta,u).
\eeq{2}
By definition the Killing vectors obey $D_A Y_B + D_B Y_A=0$. 
When it is well defined, the  total energy flux $\Delta E \equiv E(+\infty)-E(-\infty)$ is invariant under supertranslations while the angular 
momentum flux
$\Delta J_Y \equiv J_Y(+\infty)-J_Y(-\infty)$ is not, as first noticed in~\cite{pen}. 

Supertranslation-invariant definitions of the angular momentum flux for finitely radiating systems were given in~\cite{comp} and later, 
independently in~\cite{yau21a,yau21b}. A definition of the angular momentum flux in terms of bulk integrals of soft-graviton dressed 
canonical degrees of freedom was given in~\cite{jkp}. All these definitions coincide and amount to a simple prescription: replace
$N_A(u,\Theta)$ in eq.~\eqref{2} with $N_A(u,\theta) - 2 m(u,\Theta) D_A C(\Theta)$:
\beq
J^{BMS}_Y(u) = {1\over 8\pi G} \int d^2 \Theta \sqrt{h} Y^A [N_A(\theta,u)- 2 m(u,\Theta) D_A C(\Theta)].
\eeq{2a}
The superscript ``BMS'' denotes supertranslation-invariant quantities. The change in angular momentum flux implied by this prescription cannot be  reabsorbed in a local redefinition of $N_A$ because it uses the boundary graviton $C$, which is 
defined at $u=-\infty$ by
\beq
\lim_{u \rightarrow - \infty} C_{AB}(u,\Theta^A) =  -2 D_A D_B C(\Theta^A) + h_{AB}  D^2 C(\Theta^A) .
\eeq{3}
The angular momentum flux $\Delta J_Y^{BMS}$ can be written also as a 3D integral over future null infinity~\cite{jkp,hanoi}. Assuming that the
only massless field relevant to the scattering is the graviton we find 
\bea
\Delta J^{BMS}_Y &= &\frac{1}{8\pi G} \int_{-\infty}^{+\infty} du \int d^2 \Theta \sqrt{h} \, Y^A (\Theta) 
\left[ \frac{1}{2}\hat{C}_{AB}D_C N^{BC}- \frac{1}{4} D_B \left(\hat{C}^{BC}N_{CA}\right) \right]  ,
\label{4} \\ && 
\hat{C}_{AB}(u,\Theta)\equiv C_{AB}(u,\Theta) - C_{AB}(-\infty,\Theta) =\int_{-\infty}^{+\infty} du N_{AB}, \quad N_{AB}=\partial_u C_{AB}.
\eea{5}
The non-supertranslation invariant definition of $\Delta J_Y$ given in eq.~\eqref{2} is the same, except for the replacement $\hat{C}_{AB} \rightarrow C_{AB}$. 

Equation~\eqref{4} is a natural {\em but not unique} definition of the angular momentum flux. When the radiating system 
reverts back to empty space it 
defines an angular momentum that, together with translations and with the definition of boosts given in~\cite{jkp}, generates the 
Poincar\'e algebra by Poisson brackets and coincides with the usual definition of angular momentum for boosted Kerr black holes and 
Minkowski space~\cite{jkp}. Reference~\cite{yau22} proves that it is also cross-section continuous. All would be well if it weren't for a
fly in the ointment, namely the contribution of radiation reaction to gravitational scattering. It first appears at the third post-Minkowskian, $\mathcal{O}(G^3)$ order 
and has been computed by Damour in~\cite{dam} using linear response theory~\cite{bd}. The computation has been confirmed by 
several independent computations used by various groups using different methods~\cite{conf1,conf2,conf3,conf4,conf5}. The linear 
response theory requires  $\Delta J_Y$ to be $\mathcal{O}(G^2)$, but the lowest order at which the Bondi news $N_{AB}$ can be nonzero is 
$\mathcal{O}(G^2)$, so eq.~\eqref{4} implies $\Delta J^{BMS}_Y=\mathcal{O}(G^3)$! 
Clearly, the angular momentum flux computed in~\cite{dam} and derived 
by different methods in~\cite{man,heis} is not the BMS invariant one. It is instead the non BMS invariant expression computed 
using~\eqref{2} {\em in a particular BMS frame}~\cite{vv}. The special frame is defined perturbatively in $G$ in terms of the initial 
scattering data, namely the energies and momenta of the incoming particles~\cite{vv}. In that frame the boundary 
graviton $C(\Theta)$ is nonzero and is also defined in terms of particle energy and momenta. It was computed explicitly in~\cite{vv} 
and is implicitly given in~\cite{man,heis} (see also~\cite{comp2}).
Because the BMS-invariant angular momentum in~\eqref{4} precisely subtracts $C(\Theta)$, it differs from the one used 
in~\cite{dam,man,heis}. The explicit relation between the flux used by Damour, which we shall call $\Delta J_Y^{D}$, and the 
BMS invariant one, $\Delta J_Y^{BMS}$ is
\beq
\Delta J_Y^{D} = \Delta J_Y^{BMS} +\frac{1}{4\pi G} \int d^2 \Theta \sqrt{h} Y^A \Delta m(\Theta)D_A\beta(\underline{p},\Theta).
\eeq{6} 
Here $\Delta m(\Theta) = m(+\infty,\Theta) -m(-\infty,\Theta)$ while $\beta(\underline{p},\Theta)$ is the boundary graviton 
computed in~\cite{vv}.
These notations make clear that $\beta$ is a function of the $n$ initial 4-momenta of the initial particles $\underline{p}=(p_1,...,p_N)$. 
Explicitly~\cite{vv}, the metric is written as $g_{\mu\nu}=\eta_{\mu\nu} + \delta g_{\mu\nu}$, $\delta g_{\mu\nu}=\mathcal{O}(G)$, the
signature is mostly plus and
\beq
\beta(\underline{p},\Theta)= \sum_{I=1}^N 2Gm_I (n \cdot v_I) \log (-n \cdot v_I).
\eeq{7}
Here $n\cdot v\equiv n^\mu v^\mu \eta_{\mu\nu}$, $v_I^\mu$ is the 4-velocity of the $I$-th particle, $m_I$ is its mass, and 
$n^\mu=(1, n^i)$, $x^i=r n^i$, $\sum_{i=1}^3 n^i n^i =1$. Notice that 
the sign of $\beta$ in~\eqref{7} is the opposite of ref~\cite{vv} because for us $\beta$ is the boundary graviton while for~\cite{vv} it is
the supertranslation that sets the shear to zero.

The angular momentum flux $\Delta J_Y^{D}$ is BMS invariant by construction since it is computed in a specific BMS frame, 
but it is not defined in terms of the true, BMS-independent asymptotic 
degrees of freedom $N_{AB}(u,\Theta)$ and $m(u,\Theta)$.

Purpose of this paper is
to present a new formula for the {\em angular momentum}, which we shall call $J^{new}_Y(u)$, possessing the following properties: 
\begin{description}
\item{(a)} It is written purely in terms of asymptotic data.
\item{(b)} It gives a supertranslation invariant total flux $\Delta J^{new}_Y=J_Y^{new}(+\infty)-J_Y^{new}(-\infty)$ such that 
$\Delta J^{new}_Y=\Delta J_Y^{D}$ for a system of massive point particles.
\item{(c)} The $J^{new}_Y(-\infty)$ generate the rotation algebra.
\item{(d)} When the limit $u\rightarrow -\infty$  is well defined, hence
in particular for stationary spacetimes, $J^{new}_Y(-\infty)=J^{BMS}_Y(-\infty)=J_Y(-\infty)$.
\end{description}
We shall make no attempt to justify our formula from any principle. In particular, we shall not try 
to understand why the Damour formula for angular momentum flux is the correct one for radiation reaction computations. We refer 
to~\cite{vv} for a discussion on this point.

To begin with we need some identities valid for any velocity vector $v$ and $\gamma=1/\sqrt{1-v^2}$
\bea
	D_A D_B(\n) &=&-h_{AB}(\gamma+\n), \label{8.a} \\
	D_A(\n)D^A(\n) &=&-[1+(\n)^2+2\gamma \n ], \label{8.b} \\
	D_B D_A \left(\frac{1}{\n} \right) &=& \frac{h_{AB}(\gamma+\n)}{(\n)^2}+2\frac{D_A(\n)D_B(\n)}{(\n)^3}, \label{8.c} \\
	D_A D^A\left(\frac{1}{\n}\right)&=& \frac{-2}{(\n)^3}(1+\gamma \n), \label{8.d} \\
	D^2 \log(-\n) &=& -1+\frac{1}{(\n)^2}.
\eea{8}
The shear tensor obtained from eqs.~(\ref{3},\ref{7}) is~\cite{dam,vv}
\bea
	C_{AB}&=& (-2D_AD_B+h_{AB}D^2)2G\sum_{I=1}^N  m_I \ni \log(-\ni)    \nonumber\\
	&=& -G\sum_{I=1}^N m_I \left[\left(\gamma_I+\frac{1}{2}\ni+\frac{1}{2\ni}\right)h_{AB}+\frac{D_A(\ni)D_B(\ni)}{\ni}\right],
\eea{9}
therefore 
\beq
	D^A C_{AB}=-2G\sum_{I=1}^Nm_I\left(-3+\frac{1}{(\ni)^2}-\frac{\gamma_I}{\ni}\right)D_B(\ni),
\eeq{10}
and from this we can easily calculate $D^AD^BC_{AB}$,
\beq
	D^BD^A C_{AB} =-4G\sum_{I=1}^Nm_I\left[4\gamma_I+3\ni+\frac{1}{(\ni)^3}\right].
\eeq{rj1}
The first two term are essential so that the sum of all three terms in the brackets has zero monopole and dipole moments, 
as it can be checked explicitly. 

We calculate next the Bondi mass aspect. For this we need to compute $g_{uu}$ in the Bondi gauge. In Minkowski space and in the
notations of ref.~\cite{vv} the metric is given by
\beq
	g^{\mu\nu}=\eta^{\mu\nu}-4G\sum_{I=1}^N \frac{m_I (v^\mu_I v^\nu_I + \frac{1}{2}\eta^{\mu\nu})}{\Gamma_I(x)}.
\eeq{rj2}
We need to make a coordinate transformation to write the metric in Bondi gauge. Metric~\eqref{rj2} can be written in terms of 
retarded time $U=t-\rho$, radius $\rho$ and angular coordinates $\theta^A$, in which a useful expansion for the functions $\Gamma$ is
\beq
	\frac{1}{\Gamma}=\frac{H(\theta^A)}{\rho}+\frac{K(U,\theta^A)}{\rho^2}+\mathcal{O}(\frac{1}{\rho^3}),
\eeq{rj3}
where $H(\theta^A)=-1/(\n) $ and 
\beq
	K(U,\theta^A)=-\frac{u(1+\gamma \n)}{(\n)^3}+h(\theta^A).
\eeq{rj4}
We consider a coordinate transformation from the original coordinates $U,\rho,\theta^A$ to Bondi gauge coordinates $u,r,\Theta^A$ 
of the form
\beq
	u=U+\delta U,\quad
	r=\rho+\delta \rho,\quad
	\Theta^A=\theta^A+\delta \theta^A.
\eeq{rj5}
Now the mass aspect can be read from the following equation
\beq
	g_{uu}=g_{UU}\left(1-\frac{\partial \delta U}{\partial u}\right)^2+
	2g_{U\rho}\left(1-\frac{\partial \delta U}{\partial u}\right)\left(-\frac{\partial \delta \rho}{\partial u}\right).
\eeq{rj6}
We have omitted subleading terms and other terms that  do not contribute to $g_{uu}$ at order $\mathcal{O}(1/r)$ and hence do not change the mass aspect. Eq.~\eqref{rj6} can be simplified to
\beq
	\delta m=r\partial_u (\delta U +\delta \rho).
\eeq{rj7}
 To find $\delta \rho$ and $\delta U$ we use the Bondi gauge conditions and the formulas for $\delta U$, $\delta \rho$, 
 $\delta \theta^A$ given in~\cite{vv}. The result is
\bea
	\delta U&=&2G\sum_{I=1}^N m_I \Big( - \ni\log(r)-\frac{u}{r}\frac{1+\gamma_I \ni}{\ni}+\mathcal{O}(\frac{1}{r^2})\Big), \nonumber\\
	\delta \rho&=& \frac{Gu}{r} \sum_{I=1}^N\frac{m_I }{(\ni)^2}\Big((1+\gamma_I\ni)(3\ni+4\gamma_I)+ \nonumber \\
	&& \frac{1+\gamma_I\ni}{\ni}-\frac{(2+\gamma_I\ni)(1+2\gamma_I\ni+(\ni)^2)}{\ni}\Big). 
\eea{rj8} 
From these equations and after some simplifications we find that $\delta m$ is
\beq
	\delta m=G\sum_{I=1}^N m_I\Big(-\frac{1}{(\ni)^3}+\frac{2\gamma_I^2-1}{\ni}\Big).
\eeq{rj9}
The Bondi mass aspect is then 
\beq
	m(-\infty,\Theta)=-G\sum_{I=1}^N\frac{m_I}{(\ni)^3}.
\eeq{rj10}
This equation coincides as it should with eq.~(5.16) of~\cite{comp}.
The relation between mass aspect and the double divergence of the shear tensor is thus
\beq
	m(-\infty,\Theta)=\frac{1}{4}  D^A D^B C_{AB} + \mbox{ monopole and dipole terms}. 
\eeq{rj11}
{\em This equation motivates our proposal for a supertranslation-invariant angular momentum flux.}  
The key observation is that all the coefficients with
$l>1$ in the expansion of $m(u,\Theta)$  in spherical harmonics of the celestial sphere can be changed by soft radiation without
any radiative loss of energy. 
This is because the mass aspect obeys the equation
\beq
\partial_u m = \frac{1}{4} D^AD^B N_{AB} - T_{uu}.
\eeq{rj12}
When no hard radiation is present $T_{uu}=0$ so eq.~\eqref{rj12} becomes 
\beq
m(+\infty,\Theta) - m(-\infty,\Theta) =\frac{1}{4}  D^AD^B C_{AB}(+\infty,\Theta) -\frac{1}{4}  D^AD^B C_{AB}(-\infty,\Theta)
\eeq{rj13}
whose right hand side has no $l=0,1$ components by construction. 
Eq.~\eqref{rj11} shows that soft gravitational reaction can make both the shear and the $l>1$ harmonics of the mass aspect used in the
 computation of $\Delta J_Y^{D}$ vanish simultaneously without producing any hard radiation ($T_{uu}=0$) 
 so in particular without radiating out any energy.
{\em We will take this as the key hint to define the specific boundary graviton to use in the definition of an angular momentum 
obeying the properties (a)-(d) listed above.} 

The most direct and physically motivated procedure to define a preferred boundary graviton
 would have been to characterize it in terms of initial data at past null infinity $\mathcal{I}^-$ by requiring that no incoming radiation 
 crosses $\mathcal{I}^-$. To extract data at future null infinity $\mathcal{I}^+$ 
 one would need to solve the scattering problem of a matter-gravity coupled
  system. This is too tall an order so we choose instead to select the preferred boundary graviton by {\em demanding} that it
  satisfies eq.~\eqref{rj11}. This definition guarantees that eq.~\eqref{rj13} can be satisfied with $C_{AB}(+\infty,\Theta)=0$ and
  $m(+\infty,\Theta)$ equal to only a monopole plus dipole term without any $l>1$ harmonic. 
  Instead of adding a term to the BMS invariant angular momentum flux, we could have 
defined the flux $\Delta J_Y$ in a specific BMS frame. Our construction is equivalent to selecting such BMS frame by requiring that the
shear of a system whose mass aspect at $u=-\infty$ contains no harmonics higher than $l=1$ vanishes:
\beq
m(-\infty,\Theta) = c+ \sum_{m=-1}^1 c_{m}Y_{1m}(\Theta) \Rightarrow C_{AB}(-\infty,\Theta)=0.
\eeq{equiv}
Equation~\eqref{3} plus the requirement that the initial shear vanishes when the mass aspect has no dipoles higher than $l=1$ 
--i.e. that all higher moments in the mass aspect and in the shear can be radiated away without radiating energy-- imply
$\mathcal{C}_{AB}(-\infty) = (h_{AB}D^2 -2D_A D_B)\mathcal{C}$ with
$\mathcal{C}$ given by 
\beq
-(2+D^2)D^2 \mathcal{C} =D^A D^B \mathcal{C}_{AB}= 4 m(-\infty,\Theta) + \mbox{ monopole and dipole terms}.
\eeq{rj14}
Expanding both $\mathcal{C}$ and $m$ in spherical harmonics we find an explicit formula relating their angular coefficients: 
\beq
l(1-l^2)(l+2) \mathcal{C}_{lm} =4m_{lm}, \quad \forall  l>1. 
\eeq{l-harm}
We set the undefined coefficients 
$\mathcal{C}_0=\mathcal{C}_{1\pm1}=\mathcal{C}_{10}=0$. Finally, our definition of a new angular momentum is
\beq
J_Y^{new}(u) =  J_Y^{BMS}(u) + V_Y (u) ,\qquad V_Y(u)= {1\over 4\pi G} \int d^2 \Theta \sqrt{h} Y^A  m(u,\Theta)  D_A \mathcal{C}(\Theta),
\eeq{rj15}
with $\mathcal{C}$ defined in eq.~\eqref{l-harm}.

This definition obeys property (a) by construction. To check (b) we compute  
\beq
\Delta J_Y^{new}= \Delta J_Y^{BMS} + {1\over 4\pi G} \int d^2 \Theta \sqrt{h} Y^A \Delta m(\Theta)  D_A \mathcal{C}(\Theta),
\eeq{compar}
which obviously coincides with $\Delta J^D_Y$ in eq.~\eqref{6} when $\mathcal{C}=\beta$.
Since the Bondi mass aspect is a scalar under rotation of the celestial sphere the new term $V_Y(u)$ in~\eqref{rj15} transforms as a vector. Moreover, at
$u=-\infty$ $V_Y(-\infty)$ and $V_X(-\infty)$ commute for arbitrary rotations
$X,Y$, hence
$[J_Y^{BMS}(-\infty) + V_Y(-\infty), J_X^{BMS}(-\infty) + V_X(-\infty)]=
J_{[Y,X]}^{BMS}(-\infty) + 2V_{[Y,X]}(-\infty)$. So property (c) is true
if $V_Y(-\infty)=0$, that is if $V_Y(-\infty)$
satisfies property (d).
Finally, to check property (d) we notice first that 
the operator $Y^AD_A$ acts as an infinitesimal rotation on any scalar function, so it maps an $l$-th spherical harmonic into a linear 
combination of harmonics with the same $l$. Since $\mathcal{C}_l=0$ for $l=0,1$ the $l=0,1$ harmonics of $m(u,\Theta)$ do not
contribute to eq.~\eqref{rj15}. This allows us to substitute $m(-\infty)\rightarrow -(2+D^2)D^2 \mathcal{C}/4$ in~\eqref{rj15}, 
evaluate it at $u=-\infty$ and find
\beq
J_Y^{new}(-\infty) = J_Y^{BMS}(-\infty) -{1\over 16\pi G} \int d^2 \Theta \sqrt{h} Y^A  [(2+D^2)D^2 \mathcal{C}] D_A\mathcal{C}.
\eeq{rj16}
The last term vanishes as it can be proven by expanding $\mathcal{C}$ in spherical harmonics $\mathcal{C}=\sum_l C^l$ and 
using again that $Y^AD_A$ is a rotation so 
\bea
\int d^2 \Theta \sqrt{h} Y^A [(2+D^2)D^2 \mathcal{C}] D_A \mathcal{C}&=& 
\sum_l l(l^2-1)(l+2)\int d^2 \Theta \sqrt{h} Y^A \mathcal{C}^l D_A \mathcal{C}^l \nonumber \\ &=& {1\over 2} l(l^2-1)(l+2)\int d^2 \Theta 
\sqrt{h} Y^A D_A\left(\mathcal{C}^l\right)^2=0.
\eea{rj17}
In the last step we integrated by parts and used $D_AY^A=0$.

\subsection*{Acknowledgements} 
 We thank G. Comp\`ere and G. Veneziano for useful comments on the draft. M.P. would like to thank the UCLA Physics 
 Department, the Perimeter Institute and the Institute for Advanced Studies, Princeton NJ, for their kind hospitality during various 
 stages of this work. M.P.  is supported in part by NSF grant PHY-2210349.


\end{document}